\documentclass[
  journal=pasa,
  manuscript=research-paper, 
  year=2023,
  volume=31,
]{cup-journal}

\usepackage{microtype,siunitx,booktabs}
\title{Classical Novae in the ASKAP Pilot Surveys}

\author{Ashna Gulati}
\affiliation{Sydney Institute for Astronomy, School of Physics, The University of Sydney, NSW 2006, Australia}
\alsoaffiliation{ARC Centre of Excellence for Gravitational Wave Discovery (OzGrav), Hawthorn, VIC 3122, Australia}

\author{Tara Murphy}
\affiliation{Sydney Institute for Astronomy, School of Physics, The University of Sydney, NSW 2006, Australia}
\alsoaffiliation{ARC Centre of Excellence for Gravitational Wave Discovery (OzGrav), Hawthorn, VIC 3122, Australia}

\author{David L.\ Kaplan}
\affiliation{Department of Physics, University of Wisconsin-Milwaukee, P.O. Box 413, Milwaukee, WI 53201, USA}

\author{Roberto Soria}
\affiliation{Sydney Institute for Astronomy, School of Physics, The University of Sydney, NSW 2006, Australia}
\alsoaffiliation{INAF-Osservatorio Astrofisico di Torino, Strada Osservatorio 20, 10025, Pino Torinese, Italy}
\alsoaffiliation{College of Astronomy and Space Sciences, University of the Chinese Academy of Sciences, Beijing 100049, China}

\author{James K.\ Leung}
\affiliation{Sydney Institute for Astronomy, School of Physics, The University of Sydney, NSW 2006, Australia}
\alsoaffiliation{ARC Centre of Excellence for Gravitational Wave Discovery (OzGrav), Hawthorn, VIC 3122, Australia}
\alsoaffiliation{CSIRO Space and Astronomy, PO Box 76, Epping, NSW 1710, Australia}

\author{Yuanming Wang}
\affiliation{Sydney Institute for Astronomy, School of Physics, The University of Sydney, NSW 2006, Australia}
\alsoaffiliation{ARC Centre of Excellence for Gravitational Wave Discovery (OzGrav), Hawthorn, VIC 3122, Australia}
\alsoaffiliation{CSIRO Space and Astronomy, PO Box 76, Epping, NSW 1710, Australia}

\author{Joshua Pritchard}
\affiliation{Sydney Institute for Astronomy, School of Physics, The University of Sydney, NSW 2006, Australia}
\alsoaffiliation{ARC Centre of Excellence for Gravitational Wave Discovery (OzGrav), Hawthorn, VIC 3122, Australia}
\alsoaffiliation{CSIRO Space and Astronomy, PO Box 76, Epping, NSW 1710, Australia}

\author{Emil Lenc}
\affiliation{CSIRO Space and Astronomy, PO Box 76, Epping, NSW 1710, Australia}

\author{Stefan W.\ Duchesne}
\affiliation{CSIRO Space and Astronomy, PO Box 1130, Bentley WA 6102, Australia}

\author{Andrew O'Brien}
\affiliation{Department of Physics, University of Wisconsin-Milwaukee, P.O. Box 413, Milwaukee, WI 53201, USA}
\affiliation{CSIRO Space and Astronomy, PO Box 76, Epping, NSW 1710, Australia}

\doi{10.1017/pasa.2020.32}

\received {31 January 2023}
\revised  {13 March 2023}
\accepted {30 March 2023}
\published{dd Mmm YYYY}

\usepackage{hyperref}
\usepackage{adjustbox}
\usepackage{amsmath}
\usepackage{amssymb}

\usepackage{tablefootnote}

\usepackage[bottom]{footmisc}
\DeclareUnicodeCharacter{2212}{-}

\newcommand\arcsec{\mbox{$^{\prime\prime}$}}% 
\def\farcm{% 
 \mbox{.\kern -0.7ex\raisebox{.9ex}{\scriptsize$\prime$}}% 
}% 
\def\farcs{% 
 \mbox{% 
  \kern  0.13ex.% 
  \kern -0.95ex\arcsec% 
  \kern -0.1ex% 
 }% 
}% 

\begin{document}

\begin{abstract} 

We present a systematic search for radio counterparts of novae using the Australian Square Kilometer Array Pathfinder (ASKAP). Our search used the Rapid ASKAP Continuum Survey, which covered the entire sky south of declination $+41^{\circ}$ ($\sim34,000$\,square degrees) at a central frequency of 887.5 MHz, the Variables and Slow Transients Pilot Survey, which covered $\sim5,000$\,square degrees per epoch (887.5 MHz), and other ASKAP pilot surveys, which covered $\sim200-2000$\,square degrees with 2--12 hour integration times. We crossmatched radio sources found in these surveys over a two-year period, from April 2019 to August 2021, with 440 previously identified optical novae, and found radio counterparts for four novae: V5668 Sgr, V1369 Cen, YZ Ret, and RR Tel. Follow-up observations with the Australian Telescope Compact Array confirm the ejecta thinning across all observed bands with spectral analysis indicative of synchrotron emission in V1369 Cen and YZ Ret. Our light-curve fit with the Hubble Flow model yields a value of $1.65\pm 0.17 \times 10^{-4} \rm \:M_\odot$ for the mass ejected in V1369 Cen. We also derive a peak surface brightness temperature of $250\pm80$\,K for YZ Ret.  
Using Hubble Flow model simulated radio lightcurves for novae, we demonstrate that with a 5$\sigma$ sensitivity limit of 1.5 mJy in 15-min survey observations, we can detect radio emission up to a distance of 4 kpc if ejecta mass is in the range $10^{-3}\rm \:M_\odot$, and upto 1 kpc if ejecta mass is in the range $10^{-5}-10^{-3}\rm \:M_\odot$. Our study highlights ASKAP's ability to contribute to future radio observations for novae within a distance of 1 kpc hosted on white dwarfs with masses $0.4-1.25\:\rm M_\odot$ , and within a distance of 4 kpc hosted on white dwarfs with masses $0.4-1.0\:\rm M_\odot$. 

\end{abstract}

\section{INTRODUCTION}\label{s_intro}

Classical novae are eruption events on the surfaces of accreting white dwarf (WD) binaries \citep{1978ARA&A..16..171G,2012clno.book.....B,1995cvs..book.....W}. In a semi-detached binary system, Roche lobe overflow from a late type main sequence star accumulates on the WD surface, igniting a thermonuclear fusion reaction \citep[see reviews by][]{1978ApJ...226..186S,2013ApJ...777..136W}. As a result, the temperature and density at the base of the accreted envelope rise, leading to a thermonuclear runaway that causes $10^{-6}-10^{-4}\rm M_\odot$ of the accreted mass to be ejected along with heavier materials dredged up in the deeper layers of the WD \citep{1998PASP..110....3G,2007JPhG...34..431J}. The ejecta flows out at velocities ranging from 500 to 5000 $\rm km\:s^{-1}$ from the WD surface \citep{1978ARA&A..16..171G, 2020ApJ...905...62A} in a roughly spherical shell \citep{1977ApJ...217..781S} with accompanied asymmetries such as jets, rings, clumping, bipolarities \citep[e.g.,][]{2008ApJ...685L.137S,2001IAUS..205..260O,2016MNRAS.457..887W}. This eruption event enriches the interstellar medium \citep{2017MNRAS.470..401R} and has a typical energy of $10^{38}$--$10^{43}$\,$\rm ergs$. 

From weeks to years post initial optical brightening at mass ejection \citep{2011ApJS..194...28K,2017ApJ...834..196S}, the remnants of the accreted envelope on the WD surface undergo nuclear burning \citep{1977A&A....61..363S}. As the resulting luminous ionising radiation from the WD diffuses through the expanding ejecta, the peak of the spectral energy distribution of this photospheric emission from within the envelope shifts to shorter wavelengths \citep{1974ApJ...189..303G}, upto the supersoft X-ray regime through UV \citep[see reviews by][]{2021ApJS..257...49C,2020A&ARv..28....3D,2018acps.confE..57P}. Non-thermal GeV gamma ray emission is produced due to relativistic acceleration of charged particles in hadronic or leptonic scenarios that play out between the primary and hypothesised delayed mass ejection in novae \citep[see discussion by][]{2014Natur.514..339C,2015MNRAS.450.2739M,2018ApJ...852...62V,2022NatAs...6..689A}. Gamma-ray emission begins hours to days post-eruption and declines on a time-scale of days to weeks. 

Radio emission from novae is produced by either free-free thermal radiation powered by the central WD \citep{1977ApJ...217..781S,1979AJ.....84.1619H,1984ApJ...286..263T,2012clno.book.....B}, or non-thermal radiation from additional emission sources such as shocks \citep{2012BASI...40..311K}. At first, the dense ejecta stay optically thick at all radio frequencies, and we detect optically thick thermal emission from the shell's surface, resulting in a radio emission peak. As the ejecta density falls, the radio photosphere recedes through the nova shell, and we detect a mix of optically thick and optically thin thermal emission from deeper layers of the shell. When the photosphere within the envelope begins to shrink due to decreasing ejecta density, the lightcurve turns over and drops until the ejecta become optically thin throughout, resulting in relatively brighter emission at lower frequencies \citep{2012clno.book.....B}. Thermal emission traces the warm, ionised, expanding ejecta as it remains optically thick to thermal emission at radio frequencies over days to years. Non thermal emission is produced by electron acceleration at shock fronts that emit synchrotron radiation \citep{2016MNRAS.463..394V,2016MNRAS.457..887W}. Shock heated thermal gas also contributes to thermal emission, although likely in a negligible way \citep{2014MNRAS.442..713M,2023arXiv230203043S}.

Radio lightcurves of novae typically brighten after the optical peak on timescales ranging from a few weeks to a few years, with single or double peaks following from high to low frequencies \citep{2021ApJS..257...49C}. Following the spectral development of novae at high \citep{2012clno.book.....B,2014ApJ...792...57R} and low frequencies \citep{2012BASI...40..311K} shows that the radio emission from novae is consistent with multi-component emission mechanisms: with a free-free thermal component dominant at higher frequencies and a non-thermal synchrotron component, subject to variable absorption dominant at lower frequencies \citep{2020MNRAS.491.4232S,2023arXiv230110552D}. This multi-component emission mechanism is also supported in the population study of brightness temperatures in radio detected novae by \citet{2021ApJS..257...49C}. 

Radio studies of novae have helped to model the expanding ejecta and its morphology \citep[e.g.,][]{1979AJ.....84.1619H,2016PhDT........30W,1988Natur.335..235T}, yielding measurements for distances \citep[e.g., V959 Mon;][]{2015ApJ...805..136L}, shell masses \citep[e.g., T Pyx;][]{2014ApJ...785...78N} and kinetic energies \citep[e.g., Nova Aquilae;][]{1987MNRAS.228..329S}. Following the discovery of GeV gamma-ray emission from novae \citep[e.g.,][]{2010Sci...329..817A,2014Sci...345..554A,2021ApJ...910..134G,2018A&A...609A.120F}, an increasing number are being targeted at radio wavelengths in large-scale \citep[e.g.,][]{2021ApJS..257...49C,1987MNRAS.228..217B} and individual follow-up campaigns focused on novae with high optical brightness \citep[e.g., V1500 Cyg;][]{1979AJ.....84.1619H}, gamma-ray observations (\citealp[e.g., V959 Mon;][]{2014Natur.514..339C}, \citealp[V1324 Sco;][]{2018ApJ...852..108F}), and recurrent nova status \citep[e.g., U Sco;][]{2021ApJS..257...49C}.

There is still a gap between nova theory  \citep[][]{2005ApJ...623..398Y} and observation  \citep[see review by][]{2012BASI...40..293R}, particularly in terms of the observed and theoretical ejecta masses in different novae sub-types. Similarly, the mechanism of mass ejection is not completely understood, and often models assuming a single impulsive burst \citep{1986ApJ...310..222P} in a spherical outflow is too simple a geometry to explain observed  asymmetries \citep{2016MNRAS.457..887W}. Radio observations are ideal tracers of the ejecta mass because the receding radio photospere through the expanding ejecta samples the entire ejecta mass profile by interacting with it. The main reason for the uncertainty in predicting accurate ejecta masses is that the observed population of novae is biased towards the optically brightest, and, recently, gamma-ray loud novae, hence, not representative of the whole population. Furthermore, only a small subset of novae have been observed in the radio bands, mostly as multi-wavelength follow-up. To increase the accuracy of ejecta mass prediction and comparison with theoretical estimates, we need unbiased radio observations. 

The Australian Square Kilometre Array Pathfinder (ASKAP) is a widefield radio telescope that images large areas of the radio sky at frequencies between 0.8 and 1.7\,GHz \citep{2021PASA...38....9H}. The surveying capability of ASKAP allows for unbiased high cadence radio observations of novae at currently undersampled, lower radio frequencies. In this paper, we use the datasets from multi-epoch widefield surveys, such as the Variable and Slow Transients Pilot Survey \citep[VAST;][]{2021PASA...38...54M} and Rapid ASKAP Continuum survey \citep[RACS;][]{2020PASA...37...48M}, to investigate the utility of ASKAP for detecting novae. In Section \ref{s_obs}, we describe the nova sample, data acquisition, relevant methodology, and search methods. In Section \ref{s_results}, we report the four novae detected in our search. In Section \ref{s_discussion}, we discuss the novae detectability in current and future ASKAP surveys; and conclude in Section \ref{s_conc}. 

\begin{table}[t]
\centering
\caption[Survey List.]{Surveys conducted using ASKAP, included in the search for radio emission from novae in this work and the corresponding estimated integration times of the observations, $t_{int}$. Fig. \ref{Fig:Distribution} shows the survey sky coverages.}
\begin{tabular}{ll} 
\hline\hline
\text{Survey}                                               & \textbf{$t_{int}$}  \\ 
\hline 
  \begin{tabular}[c]{@{}l@{}}Variable and Slow Transients  Survey (VAST)   \end{tabular}$^{a,i}$ & $12$ min  \\
Rapid ASKAP Continuum survey (RACS)$^b$ &  $15$ min   \\
 \begin{tabular}[c]{@{}l@{}}Deep Investigation of Neutral Gas Origins (DINGO)$^c$ \end{tabular}        & $\sim8$ hrs                                                                                                                                                                  \\
Evolutionary Map of the Universe (EMU)$^d$                        & $5-10$ hrs                                                                                                                                                                                \\
    \begin{tabular}[c]{@{}l@{}}First Large Absorption Survey~in HI (FLASH)  \end{tabular}$^e$              & $\sim2$ hrs                                                                                                                                                                     \\
Galactic ASKAP survey (GASKAP)$^f$                                & \begin{tabular}[c]{@{}l@{}}$\sim12$ hrs \end{tabular}                      \\
 \begin{tabular}[c]{@{}l@{}}Polarization Sky Survey of the Universe's \\ Magnetism (POSSUM)$^g$ \end{tabular} & $\sim10$ hrs                                                                                                                                                                 \\
 \begin{tabular}[c]{@{}l@{}}Widefield ASKAP L-band Legacy All-sky \\ Blind surveY (WALLABY)$^h$ \end{tabular} & $\sim8$ hrs                \footnotetext{ $^a$\citet{2021PASA...38...54M} $^b$\citet{2020PASA...37...48M}  $^c$\citet{2009pra..confE..15M} $^d$\citet{2021PASA...38...46N}  $^e$\citet{2021arXiv211000469A}  $^f$\citet{2013PASA...30....3D} $^g$\citet{2010AAS...21547013G} $^h$\citet{2020Ap&SS.365..118K}\\
 $^i$\text{15 observation epochs from the VAST survey have been included in this search.}}  

\\
\hline
\end{tabular}
\label{Surveys}
\end{table}

\section{DATA AND METHODOLOGY}\label{s_obs}

\subsection{ASKAP data}

\begin{figure*}[t]
\begin{center}
\includegraphics[width=1\textwidth]{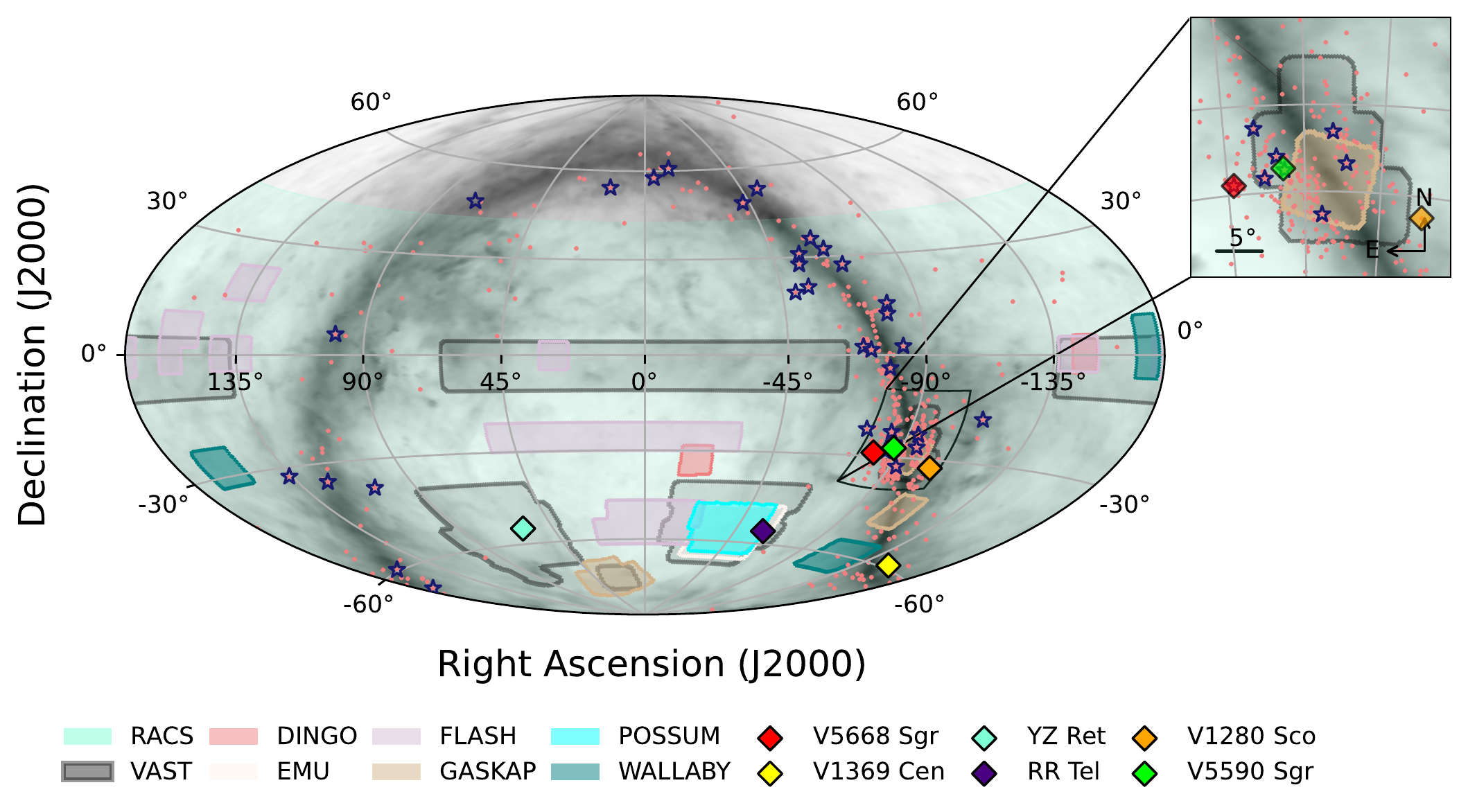}
\vspace{-0.7cm}
\end{center}
\caption[Galactic nova distribution]{Positions of novae in both samples in equatorial coordinates in relation to the ASKAP survey footprints. Circle markers show the distribution of Sample 1. Star markers show the distribution of Sample 2. Diamond markers show the positions of novae having radio emission detected at their positions, labeled by source name. Filled-in background regions depict the ASKAP pilot surveys searched for radio emission from novae, as labeled. The gray-shaded background shows the Milky Way dust distribution map from the \citet{2014A&A...571A..11P}, where the light to dark shading corresponds to extinction magnitudes, $A_V$, ranging from  $3.61\times10^{-3}$ to $1.36\times10^{3}$\,mag respectively.}
\label{Fig:Distribution}
\end{figure*}

ASKAP is a widefield radio survey instrument made up of an array of 36 prime focus antennas, located at the CSIRO Murchison Radio-astronomy Observatory in Western Australia \citep{2007PASA...24..174J,hotan21}.
Over 2020 -- 2022 a number of pilot surveys were carried out with ASKAP, and we have used these as the basis of the work presented here. From 2023, these surveys will continue as ASKAP commences full operations.

We searched for novae radio counterparts in ASKAP data from RACS, VAST, and other surveys listed in Table \ref{Surveys}. Except for RACS, all surveys listed in Table \ref{Surveys} have been in the pilot phase so far, and will be continued and expanded in ASKAP's full surveying mode starting in 2023. 
RACS covers the entire sky south of declination $\sim40^{\circ}$ \citep{2020PASA...37...48M}. In this search, we used the first epoch of RACS at 0.888\,GHz, RACS-low, conducted in 2019 \citep[declination of $-80^{\circ}$  to $41^{\circ}$;][]{2020PASA...37...48M}. We also used the second epoch of RACS at 1.367\,GHz, RACS-mid, conducted in 2021 (declination of $-90^{\circ}$ to $49^{\circ}$; Duchesne et al. {\it submitted}) to get an additional measurement for two of our candidates -- V1369 Cen and YZ Ret. The VAST project conducted $\sim162$\,hours of observations in its pilot survey phase. These were spread over multiple epochs, with the goal of detecting transient astronomical phenomena on timescales ranging from 10\,seconds to several years \citep{murphy13}. We used data from 13 epochs of the VAST pilot survey Phase 1 (VAST-P1) and two epochs of the Phase 2 (VAST-P2 (low)) observations at 0.888\,GHz from August 2019 to August 2021 \citep{2021PASA...38...54M}. 
The VAST and RACS surveys have integration times of 12-15 minutes over a total field of view of $\sim$5,000 $\rm deg^2$ and $\sim$34,000 $\rm deg^2$ respectively. All the other pilot surveys listed in Table~\ref{Surveys} have much longer integration times, $t_{int}$, for each field over smaller regions.

\subsection{Sample selection}
\label{S_2}

We constructed two samples of novae to search for in the ASKAP surveys. 
For our first sample, we selected novae from the list maintained by Bill Gray under Project Pluto\footnote{\url{https://projectpluto.com/galnovae/galnovae.htm}}. We included classical novae; objects that were classified as fast novae (NA), slow novae (NB), extremely slow novae (NC), recurrent novae (NR), and those that have not yet been assigned a confirmed classification (for example, NC/ZAND, UG/N?). We excluded objects classified as symbiotic Z Andromedae systems (Z And class), U Gem variables or dwarf novae, X-ray nova systems and variable star types -- Mira, UV Ceti, FU Ori and S Dor class.  
As of July 2021, there are 470 identified classical novae, recurrent novae, and previously unclassified candidate novae systems discovered at optical wavelengths in the Milky Way. Four hundred and forty of those 470 sources occured within the region covered by one or more ASKAP surveys. These 440 objects constitute Sample~1 of our targeted search. 

\begin{figure*}[]
    
    \centering
     \includegraphics[width=0.495\textwidth,trim={0 0 0 0},clip]{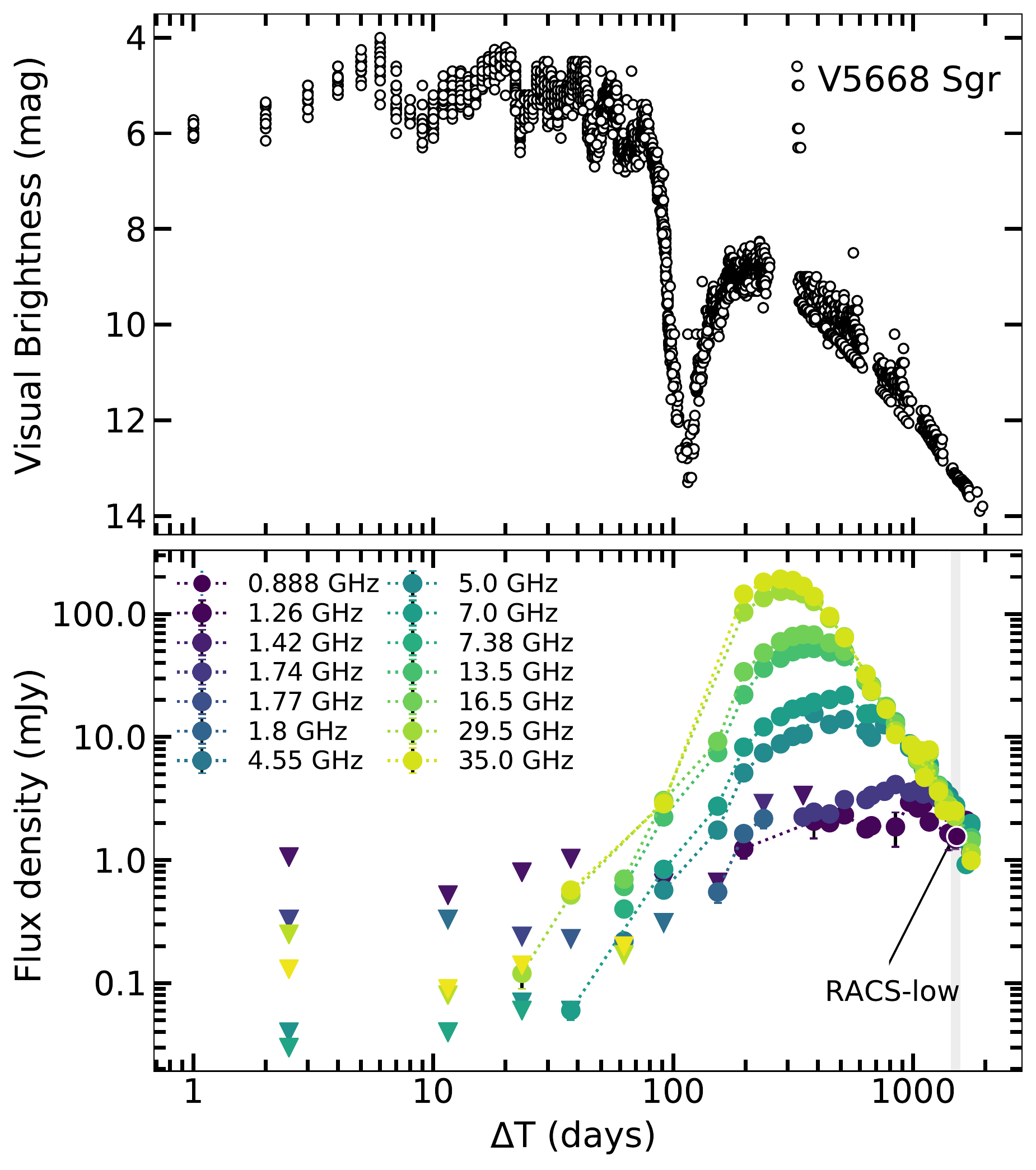}
     \includegraphics[width=0.495\textwidth,trim={0 0 0 0},clip]{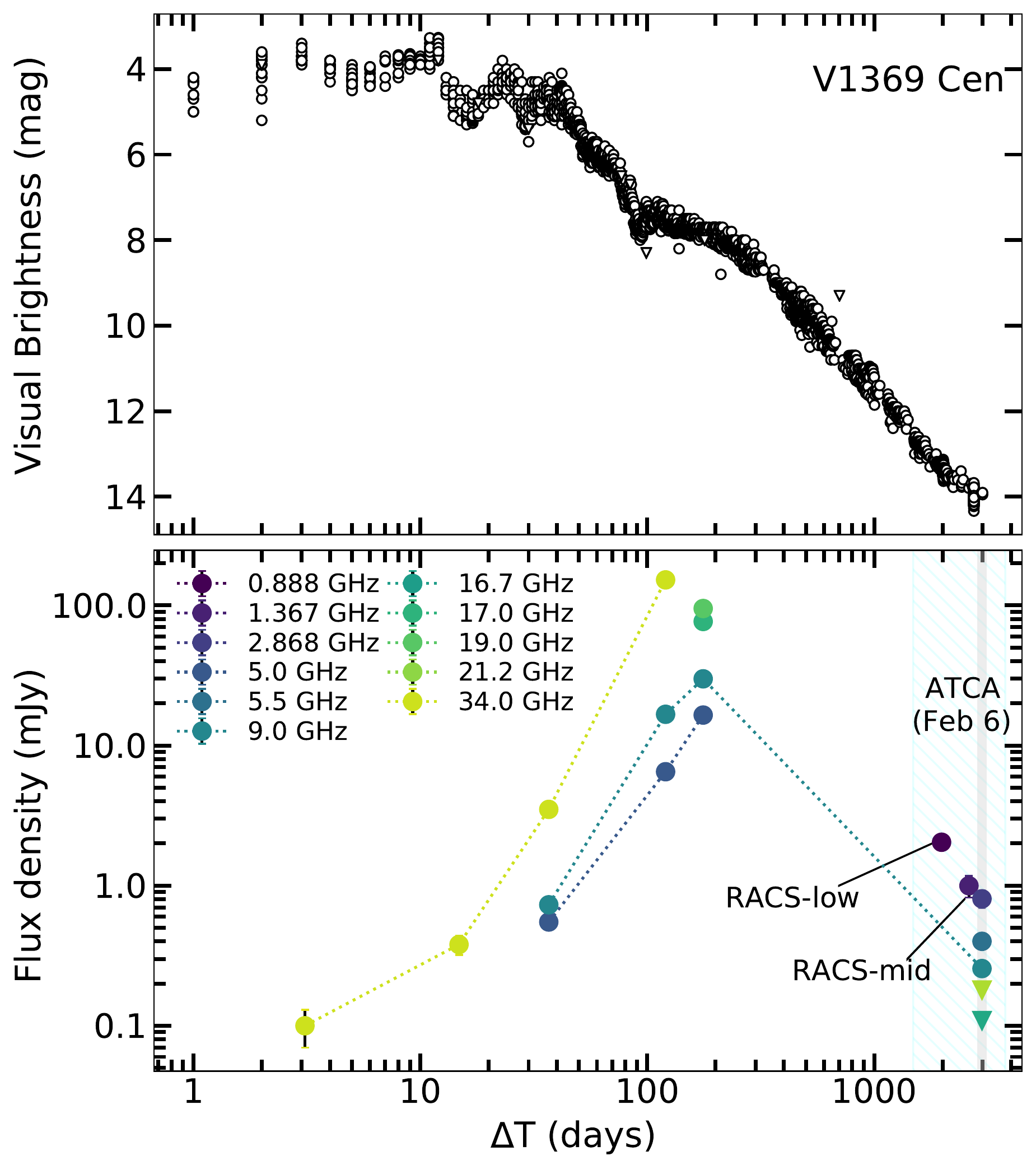}
     \\

     \includegraphics[width=0.495\textwidth,trim={0 0 0 0},clip]{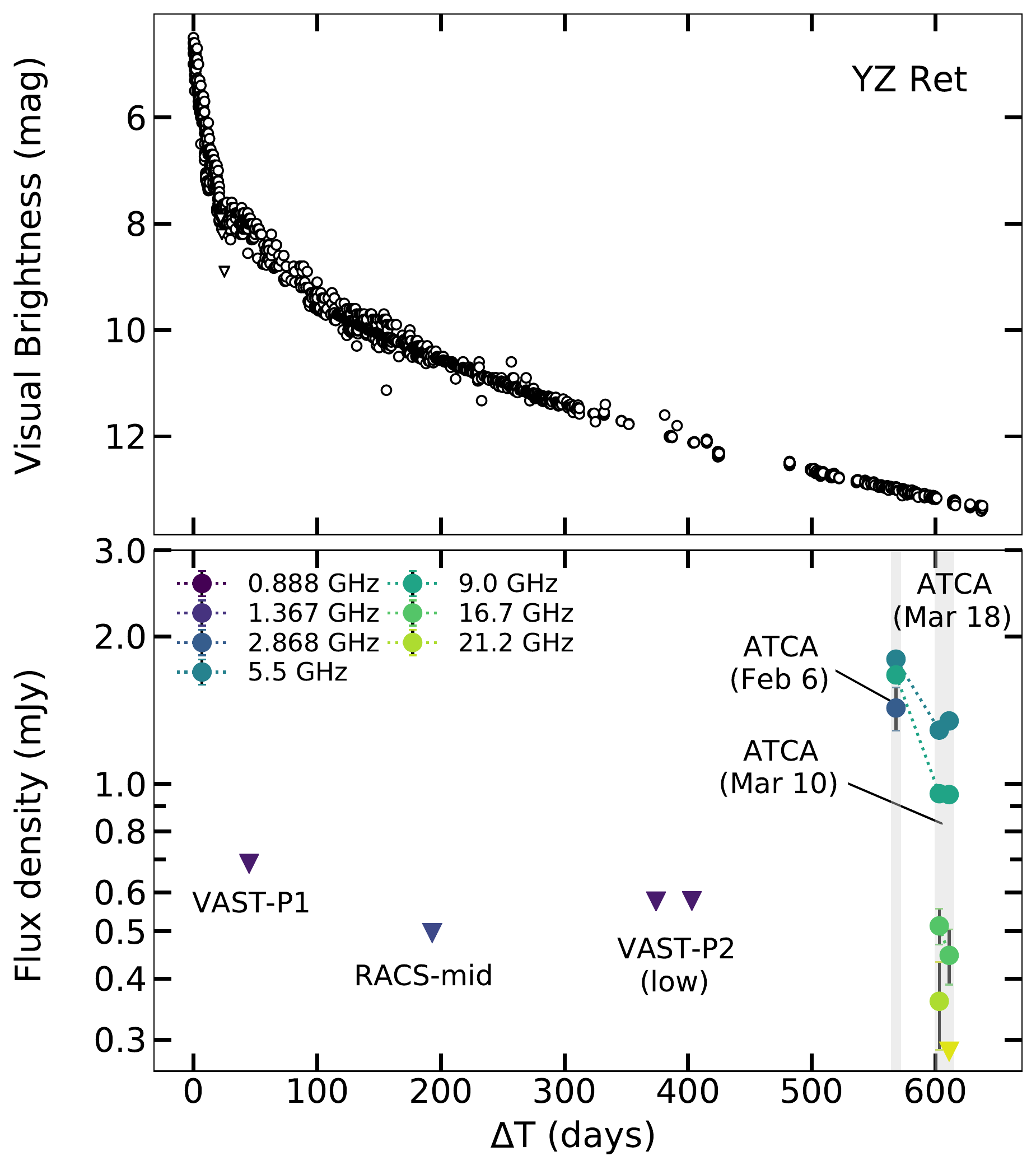}
     \includegraphics[width=0.495\textwidth,trim={0 0 0 0},clip]{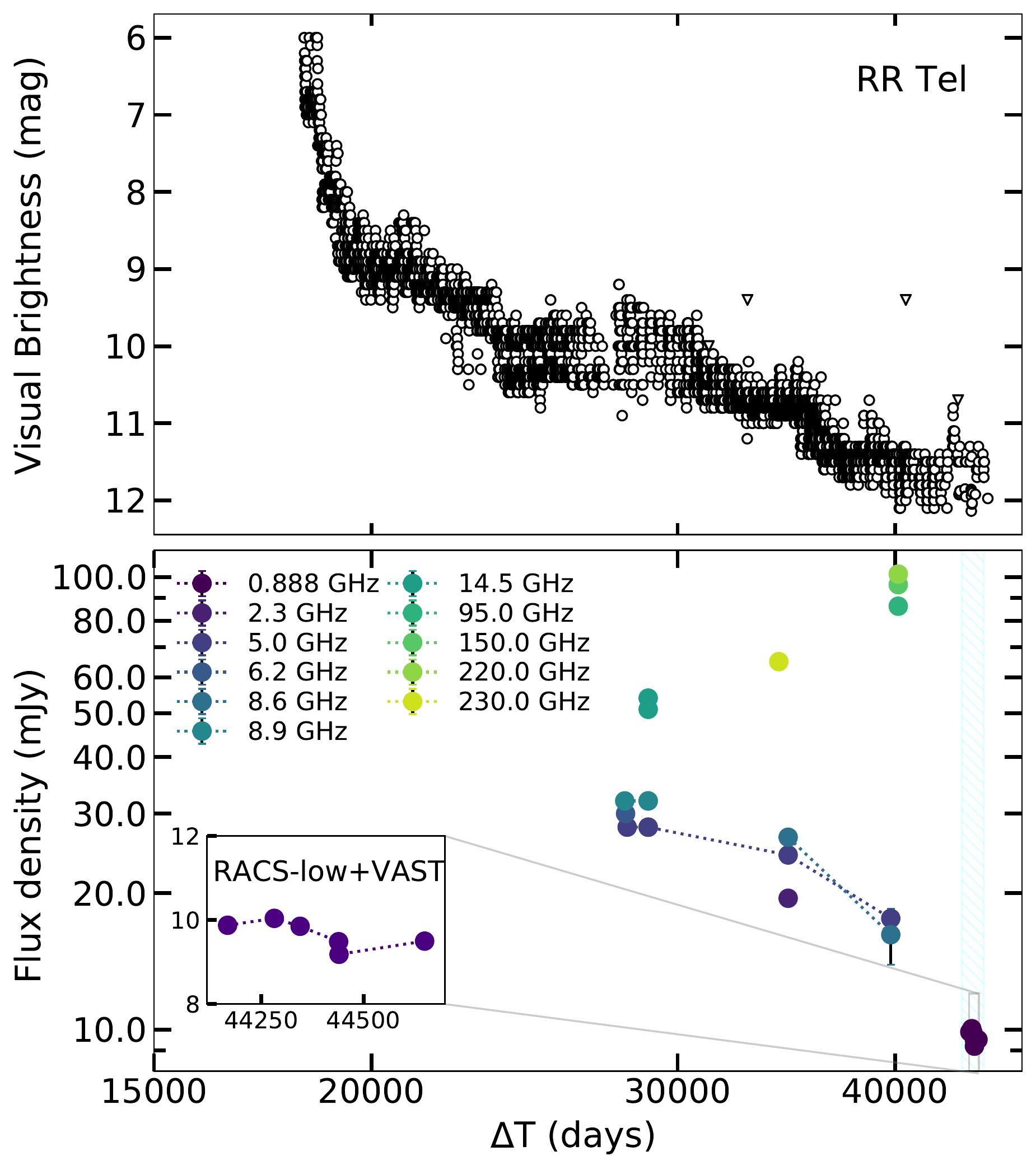}\\
     
\caption[]{Radio lightcurves for V5668 Sgr, V1369 Cen, YZ Ret, and RR Tel. For each source, the top panels show the optical light curve created using V -band or visual AAVSO measurements  \citep{Kafka_2021}. The bottom panel for: V5668 Sgr shows the radio lightcurve created using ASKAP data observed with the RACS survey ($\alpha_{1.26-1.74\rm\,GHz} = 1.65\pm0.55$; $\alpha_{1.74-7.0\rm\,GHz} = 0.07\pm0.02$ for the grey stripe) and observations from \citet{2022yCat..22570049C}; V1369 Cen shows the radio lightcurve created using  ASKAP and ATCA observations from this work highlighted in the blue striped region, and  observations from \citet{2022yCat..22570049C} ($\alpha_{2.9-9.0\rm\,GHz}= -1.00\pm0.05$ for the grey stripe); YZ Ret shows the radio lightcurve (time in linear space) created using ASKAP and ATCA observations from this work -- $\alpha_{1.7-5.5\rm\,GHz} = 0.85\pm0.32$; $\alpha_{5.5-9.0\rm\,GHz} = -0.15$ (2022 Feb 6); $\alpha_{5.5-21.2\rm\,GHz}=-0.94\pm0.10$ (2022 Mar 10) and $\alpha_{5.5-16.7\rm\,GHz}= -1.00\pm0.15$ (2022 Mar 18); and RR Tel shows the radio lightcurve created using ASKAP data observed with VAST and RACS surveys and observations from \citet{1977ApJ...211..547S,1978MNRAS.184..893W,1982MNRAS.198..321P,1995MNRAS.273..517I,2021ApJ...911...30D,2003MNRAS.342.1117M}. Downward triangle markers are 3$\sigma$ detection limits.}%
\label{lightcurves}%
\end{figure*}

For our second sample we used the dataset of radio observations compiled by \citet{2021ApJS..257...49C,2022yCat..22570049C}. This consists of 36 classical or not yet confirmed recurrent novae discovered after 1980, and well-sampled at radio wavelengths. Most of the data is from published and archival unpublished Very Large Array (VLA) datasets, as well as new observations made with the Australian Telescope Compact Array (ATCA). Fig. \ref{Fig:Distribution} shows the positions of novae from Samples 1 and 2 in relation to the ASKAP pilot survey footprints\footnote{ASKAP pilot survey footprints available at: \url{https://github.com/askap-vast/askap_surveys}}. For each nova in the samples above, we recorded the discovery date from literature as $t_0$ (refer to Table \ref{table:detections}), and plot all times relative to that $t_0$ as $\Delta T$ in Fig. \ref{lightcurves}. We additionally recorded the distances, optical variability type and previous multi-wavelength observations from relevant literature for each nova in Section \ref{Notes_on_detected_novae}.

\begin{table*}[ht!]
\sffamily
\caption[Survey List.]{ATCA and ASKAP observations for ASKAP-detected novae in this targeted search. Column 1 gives the GCVS ID of the nova detected; Column 2 gives the GCVS class \citep{1981PASP...93..165D} or variability type of the novae; Columns 3 and 4 give the optical coordinates of the nova; Column 5 gives the nova discovery date, $t_0$, in the optical band; Column 6 gives the radio observation date; Column 7 gives the time of the observations in days post-discovery in optical, $\Delta T$; Column 8 gives the radio telescope or survey used for the observations; Column 9 gives the central frequency of the observations, $\nu$; and Column 10 gives the catalogued integrated flux, $S_{int}$, for detections, and $3\sigma$ limits for non-detections.}
\begin{adjustbox}{max width=\textwidth}
\begin{tabular}{rrrrlrrrrr}
\hline
\hline
GCVS ID & GCVS class & \begin{tabular}[c]{@{}c@{}}RA\\ (J2000)\end{tabular} & \begin{tabular}[c]{@{}c@{}}Dec\\ (J2000)\end{tabular} & \begin{tabular}[c]{@{}c@{}}$t_0$\\ (UTC)\end{tabular} & \begin{tabular}[c]{@{}c@{}}Observation Date\\ (UTC)\end{tabular} & \begin{tabular}[c]{@{}c@{}}$\Delta T$\\ (days)\end{tabular} & \begin{tabular}[c]{@{}c@{}}Telescope/\\ Survey\end{tabular} & \begin{tabular}[c]{@{}c@{}}$\nu$\\ (GHz)\end{tabular} & \begin{tabular}[c]{@{}c@{}}$S_{int}$\\ (mJy)\end{tabular} \\ \hline
V5668 Sgr & N & 18:36:56.83 & $-$28:55:39.98 & 2015 Mar 15 & 2019-04-28 & 1505 & ASKAP/RACS-low & 0.9 & $1.56\pm0.06$ \\ \hline
V1369 Cen & NA & 13:54:45.35 & $-$59:09:04.09 & 2013 Dec 2 & 2019-05-06 & 1981 & ASKAP/RACS-low & 0.9 & $2.04\pm0.04$ \\ 
 &  &  &  & & 2021-02-01 & 2615 & ASKAP/RACS-mid & 1.4 & $0.99\pm0.18$ \\
 &  &  &  &  &  2022-02-06 & 2985 & ATCA & 2.9 & $0.81\pm0.12$ \\
 &  &  &  &  &  &  & &   5.5 & $0.40\pm0.05$ \\
 &  &  &  &  &  &  & &   9.0 & $0.26\pm0.03$ \\
 &  &  &  &  &  &  & &   16.7 & $< 0.07$ \\
 &  &  &  &  &  &  & & 21.2 & $< 0.12$ \\

\hline
 YZ Ret &  & 03:58:29.55 & $-$54:46:41.23 & 2020 July 15 & 2020-08-29 & 45 &  ASKAP/VAST-P1 &  0.9 & $<0.69$ \\
 & & & &  & 2021-01-24 & 193 & ASKAP/RACS-mid & 1.4 & $<0.51$\\
 
& & & & & 2021-07-24 & 374  & ASKAP/VAST-P2 (low) & 0.9 & $< 0.58$ \\
 &  & & & & 2021-08-22 & 403 & & 0.9 & $<0.58 $   \\
 
 & & & & & 2022-02-06 & 568 & ATCA & 2.9 & $1.43\pm0.15$ \\
 &  &  &  &  &  &  & &   5.5 & $1.80 \pm0.04$ \\
 &  &  &  &  &  &  & &   9.0 & $1.67\pm 0.03$ \\
 &  &  &  &  &  2022-03-10 & 603 &  & 5.5 & $1.29\pm 0.04$ \\
 &  &  &  &  &  &  & &   9.0 & $0.96 \pm0.04$ \\
 &  &  &  &  &  &  & &   16.7 & $0.51\pm 0.04$ \\
 &  &  &  &  &  &  & &   21.2 & $0.36 \pm0.07$ \\
 &  &  &  &  &  2022-03-18 & 611 &  & 5.5 & $1.35 \pm 0.04$ \\
 &  &  &  &  &  &  & &   9.0 & $0.95 \pm0.03$ \\
 &  &  &  &  &  &  & &   16.7 & $0.45 \pm 0.06$ \\
&  &  &  &  &  &  &  & 21.2 & $<0.28$ \\ 

\hline
RR Tel & NC/ZAND & 20:04:18.54 & $-$55:43:33.15 & 1898 & 2019-05-06 & 44168 & ASKAP/RACS-low & 0.9 & $9.88 \pm0.08$ \\
 &  &  &  &  &  2019-08-28 & 44282 & ASKAP/VAST-P1 & 0.9 & $10.04\pm 0.17$ \\
 &  &  &  &  &  2019-10-30 & 44345 &  &  & $9.85 \pm 0.05$ \\
 &  &  &  &  &  2020-02-01 & 44439 &  &  & $9.49 \pm 0.06$ \\
 &  &  &  &  &  2020-02-02 & 44440 &  &  & $9.19 \pm 0.00$ \\
 &  &  &  &  &  2020-08-29 & 44649 &  &  & $9.50 \pm 0.08$ \\
 &  &  &  &  &  2021-07-22 & 44976 & ASKAP/VAST-P2 (low) & 0.9 & $9.19 \pm 0.34$ \\
 &  &  &  &  &  2021-08-24 & 45009 &  &  & $9.29 \pm 0.34$ \\ \hline
\end{tabular}
\end{adjustbox}
\label{table:detections}
\end{table*}

\subsection{Search method}
\label{Search method}

We searched for radio counterparts by crossmatching (search radius \ang{;;5}) the proper motion corrected optical positions from SIMBAD (corrected to date: 2020~July~1) for novae in the two samples with sources catalogued in the surveys in Table \ref{Surveys}. We extracted radio data, images, and flux density estimations at the optical nova positions for VAST using \texttt{VAST Tools}\footnote{\texttt{VAST Tools} (\url{https://www.vast-survey.org/vast-tools/}) is a python module customized to interact with and obtain the VAST Pilot Survey data \citep{2021PASA...38...54M}.}. We obtained the source finder output catalogues for the other surveys from the CSIRO ASKAP Science Data Archive (CASDA)\footnote{\url{https://research.csiro.au/casda/}}. 
The signal to noise ratio for detections in the main catalogues is $5\sigma$. We also searched for radio counterparts at a lower signal to noise ratio of $3\sigma$ by measuring the flux density and background noise at each nova position in the extracted image cutouts.

We examined the positional offsets between our candidate radio counterparts and their matched proper-motion-corrected optical progenitor positions to determine whether our potential candidates were genuine counterparts. We also looked for previous radio emission in surveys such as the Very Large Array Sky Survey \citep[][]{2021ApJS..255...30G,lacy20}, the National Radio Astronomy Observatory VLA Sky Survey \citep[][]{1998AJ....115.1693C}, the Sydney University Molonglo Sky Survey \citep[][]{2003MNRAS.342.1117M},  the second epoch Molonglo Galactic Plane Survey \citep[][]{2007MNRAS.382..382M}, The GMRT 150 MHz All-Sky Radio Survey alternative data release 1 \citep[][]{2017A&A...598A..78I} and The Galactic and Extra-Galactic All-Sky MWA \citep[][]{2017MNRAS.464.1146H}. Where radio data were available, we checked to see if the potential candidate's radio flux density in the ASKAP observations was consistent with expected nova evolution (i) in time, i.e., radio lightcurves from earlier radio observations, and (ii) in frequency, i.e., spectral energy distributions from concurrent multi-frequency observations per the evolution of the radio spectral index, $\alpha$ (defined as $S_{\nu} \propto \nu^{\alpha}$). 
\citet{2012clno.book.....B} suggests a spectral index, $\alpha$, approaching 2 for rising optically thick thermal bremsstrahlung and $\alpha\sim -0.1$ for declining optically thin thermal emission. A steep spectrum of $\alpha\sim-0.75$ is typical for optically thin synchrotron emission (observed in for example, V3885 Sgr; \citet{2011MNRAS.418L.129K}, GK Persei; \citet{1984ApJ...281L..33R}, RS Oph; \citet{2007ApJ...667L.171K} and V1723 Aql; \citet{2016MNRAS.457..887W}). The late-time transition from rising to a falling spectrum is characterised by one or two spectral breaks \citep{2016MNRAS.457..887W}.

We quantified the false-positive rate for chance alignment between optical nova positions and unrelated sources using a Monte Carlo analysis. We took the sample of 470 novae described in \S\ref{S_2} and offset their positions by a random angle and a random distance between \ang{;1;}-\ang{;10;} and calculated the number of detections within a crossmatch radius of $\ang{;;5}$ for each random offset of the original nova distribution in the RACS survey. Unlike the original distribution, each of the 100 offset distributions with 470 sources produced $\leq1$ match (22 distributions had 0 matches, 78 distributions had 1 match) to the 2.1 million radio sources, and no distribution had five positive matches in RACS. This implies a false positive rate of $4.8 \times 10^{-4}$, indicating that the majority of the novae we detected in RACS are real.  

\section{RESULTS}\label{s_results}

We found candidate radio counterparts for six novae from Sample 1. All of these were detected in RACS, and two (RR Tel and V5590 Sgr) were also detected in VAST. Two of these six novae (V1369 Cen and V5668 Sgr) are also in Sample 2. We found no detections in the WALLABY, GASKAP, FLASH, DINGO, EMU, and POSSUM surveys. 

After further investigation we confirmed radio counterparts for four objects: the extremely slow symbiotic nova RR Tel and the classical novae V1369 Cen, V5668 Sgr, and YZ Ret. All confirmed radio counterparts were spatially consistent with their proper-motion-corrected optical progenitor positions within astrometric errors. The conservative astrometric error for a weak source ($6\sigma$) in RACS is of the order $\sim\ang{;;2}.5$ \citep[see][]{2020PASA...37...48M}. Two of the counterparts --- those for V1280 Sco and V5590 Sgr --- are uncertain, pending further multi-frequency radio observations. These are discussed in \ref{Ruled-out Nova Counterparts}. 

Our ASKAP and ATCA observations are summarised in Table \ref{table:detections}, and discussed below.

\subsection{Follow-up ATCA observations}

We observed the radio counterpart of ASKAP-detected nova V1369 Cen on 2022 February 6, with the ATCA (Project code C3363; PI T. Murphy), measuring the flux density at a wide range of frequencies to determine its evolution phase. 
In addition, based on a tentative $\geq3\sigma$ detection (Lenc, private communication) in preliminary RACS-mid data, we observed YZ Ret on 2022 February 6, March 10, and March 18 to confirm the ASKAP detection. All our observations used 2048 MHz-wide bands centered on 2.1, 5.5, 9.0, 16.7, and 21.2\,GHz, except the last epoch that had a zoom mode observation bandwidth of 58\,MHz in the L-band.

We reduced the visibility data using standard routines in \texttt{MIRIAD} \citep{1995ASPC...77..433S}. We used a combination of manual and automatic RFI flagging before calibration, conducted with \texttt{MIRIAD} tasks \mbox{} \texttt{uvflag} and \texttt{pgflag}, respectively. We used PKS~1934$-$63 to determine the bandpass response for all frequency bands in all epochs, except the 16.7/21.2 GHz bands. For these two exceptions, we used  PKS~0727$-$115 as the bandpass calibrator for the first epoch and PKS~1921$-$293 for the second epoch. We used PKS~1934$-$63 to calibrate the flux density scale for all epochs and frequency bands for both sources. For YZ Ret, we used PKS~0334$-$546 to calibrate the time-variable complex gains for all epochs and frequency bands. For V1369 Cen, we used IVS~B1338$-$582 to calibrate the gains for the 2.9/5.5/9.0-GHz bands and IVS~B1325$-$558  for the 16.7/21.2-GHz bands. After calibration, we inverted the visibilities using a robust weighting of 0.5 and then used the CLEAN algorithm \citep{1980A&A....89..377C} with 250 iterations to the target source field using standard \texttt{MIRIAD} tasks \texttt{INVERT}, \texttt{CLEAN} and \texttt{RESTOR} to obtain the final images.

\subsection{Notes on Individual Objects}
\label{Notes_on_detected_novae}

 We discuss each source briefly below. Fig. \ref{lightcurves}, shows the lightcurve for each object, with all available radio data. 

 \subsubsection{V5668 Sgr}

V5668 Sgr is a slow-evolving dust-forming classical nova \citep{2018A&A...611A...3H,2021ApJS..257...49C} discovered by J.~Seach on 2015 March 15.634~UT, at a distance of $2.8\pm0.5$\,kpc \citep{2021ApJ...910..134G}. It is well-sampled at centimetre radio wavelengths with the VLA and ATCA \citep{2021ApJS..257...49C}, and has near-simultaneous detections in the RACS survey with ASKAP (see Fig. \ref{lightcurves} \textit{top panel left image}). Ejecta morphology and dust formation have been studied for V5668 Sgr at mid- \citep{2015ATel.7862....1G} and near-infrared \citep{2016MNRAS.455L.109B}, X-ray  \citep{2018ApJ...858...78G} and millimetre wavelengths  \citep{2018MNRAS.480L..54D}. The radio counterpart of V5668 Sgr in RACS is offset by \ang{;;2}.8 from its optical position.

We measured a 0.888\,GHz flux density of $1.56\pm0.06$\,mJy for V5668 Sgr (1505 days post-discovery in optical). This value agrees with the unbroken power-law extrapolation of flux densities measured in near-simultaneous, higher frequency radio observations with the ATCA at 1500.5 days by \citet{2021ApJS..257...49C}. For the near-simultaneous multi-frequency ASKAP and ATCA detections at this epoch, the peak radio flux density is somewhere between 1.26 and 5 GHz ($\alpha_{1.26-1.74\rm\,GHz} = 1.65\pm0.55$; $\alpha_{1.74-7.0\rm\,GHz} = 0.07\pm0.02$), which indicates that the emission is a mix of optically thick and optically thin thermal emission and at the point of spectral turnover optical depth, $\tau=1$. This characteristic indicates that the classical nova V5668 Sgr was in its declining radio evolution phase at the time of our observations.

\subsubsection{V1369 Cen}
V1369 Cen was discovered in an eruption on 2013 December 2.69 UT by J.~Seach at a distance of $1.0\pm0.1$\,kpc \citep{2021ApJ...910..134G}. It was detected with the ATCA between 2013 Dec 5 and 2014 May 27 in multi-frequency observations (5.5 - 34\,GHz) by \citet{2014ATel.6058....1B, 2021ApJS..257...49C}. It is characterised as a fast radio classical nova by \citet{2021ApJS..257...49C}. Despite  multiwavelength studies of the object at near-IR \citep{2021A&A...649A..28M}, UV \citep{2014ATel.6413....1S}, gamma-ray \citep{2016ApJ...826..142C}, and X-ray wavelengths \citep{2021ApJ...910..134G}, radio observations of V1369 Cen around its thermal maximum are limited. The radio counterpart of V1369 Cen in RACS is offset by \ang{;;0}.9 from its optical position. 

We measured a 0.888\,GHz flux density of $2.04\pm0.04$\,mJy for V1369 Cen. We also measured a $5\sigma$ detection for the source in the RACS-mid survey (1.36\,GHz). For our 2022 Feb 6 ATCA follow-up observation, we measured a $\geq10\sigma$ detection at 2.1/5.5/9\,GHz and non-detections at 16.7/21.2\,GHz \citep{2022ATel15310....1G}. In this epoch, the steep spectral index, $\alpha_{2.9-9.0\rm\,GHz}= -1.00\pm0.05$, is indicative of synchrotron emission. Peaking of synchrotron radiation, dominant at lower frequencies due to decreasing free-free emission, suggests that the ejecta are optically thin across the observed bands. These characteristics indicate that classical nova V1369 Cen was in its declining radio evolution phase at the time of our observations (see Fig. \ref{lightcurves} \textit{top panel right image}).

\begin{figure}[t]
     \includegraphics[width=1\textwidth,trim={0cm 0cm 0cm 0cm},clip]{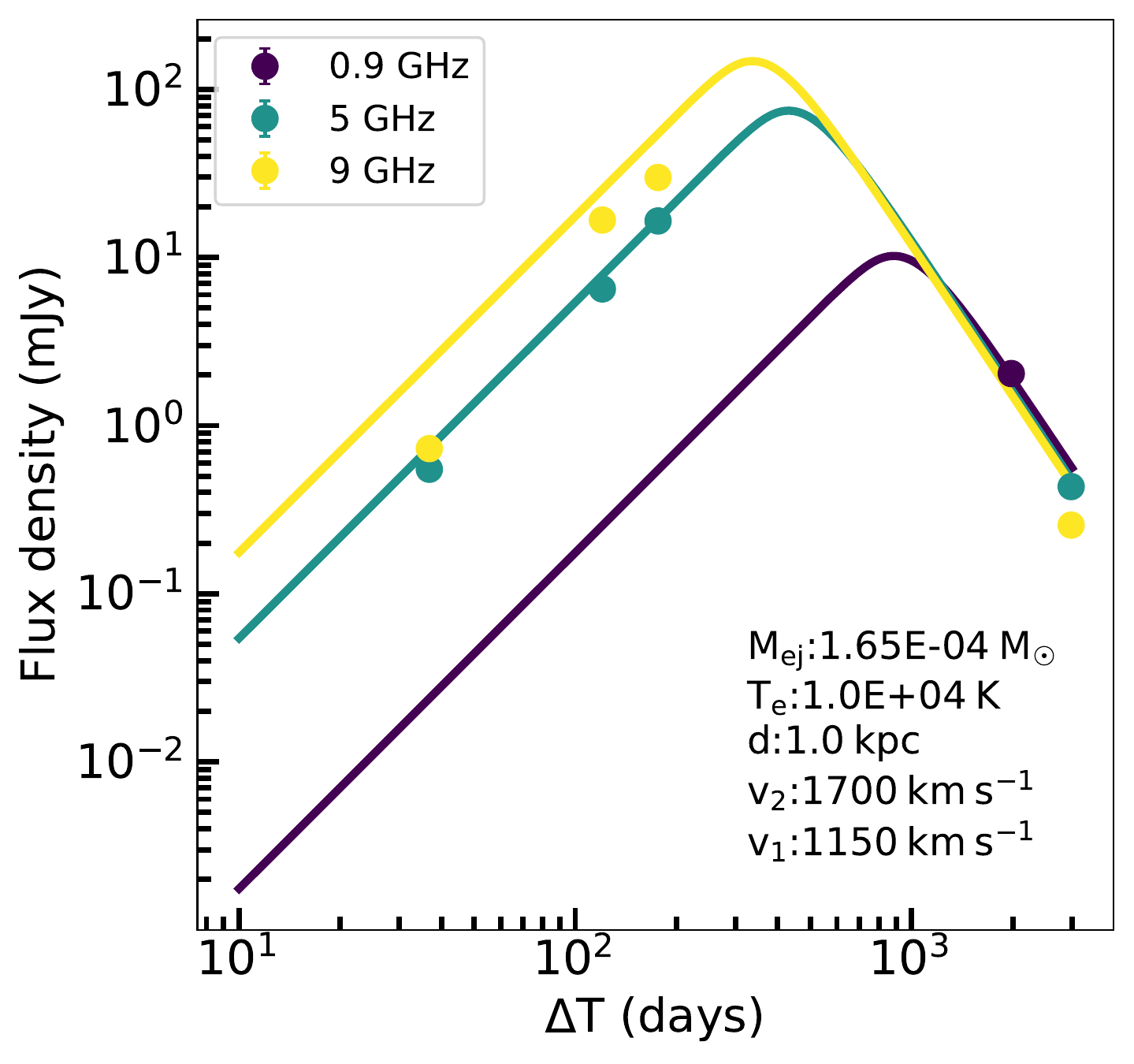}   
\caption{Hubble Flow model fit to V1369 Cen radio detections. Minimization of $\chi^2$ between 0.9, 5 and 9\,GHz radio detections, given by filled-in circles, and models with ejecta masses ranging between $5\times10^{-6}\: \rm M_\odot$ and $1\times10^{-3}\: \rm M_\odot$, results in the fit in the teal and yellow solid line fits respectively.}%
\label{V1369_hubble_flow_fit}
\end{figure}

We fit a Hubble Flow model\footnote{Hubble Flow model available at: \url{https://github.com/askap-vast/source_models}} \citep[described in ][]{1979AJ.....84.1619H,1996ASPC...93..174H,1977ApJ...217..781S,1980AJ.....85..283S} to multi-frequency radio detections (0.9, 5 and 9\,GHz) for V1369 Cen obtained in this work as well as previous works via $\chi^2$ \footnote{$\chi^2 = \sum^{n}_{i=1} \bigg(\frac{S_{i} - \widehat{S_{i}}}{\sigma_{i}} \bigg)^2$, where $S_i$ is the measured flux density, $\widehat{S_{i}}$ is the model-fitted flux density at the same observation epoch, and $\sigma_i$ is the measurement error.} minimisation between the data and models with ejecta masses ranging between $5\times10^{-6}\: \rm M_\odot$ and $1\times10^{-3}\: \rm M_\odot$ (see Fig. \ref{V1369_hubble_flow_fit}). We assumed that the mass was ejected at the time of the optical discovery. We used model parameters previously derived in literature for V1369 Cen, such as velocities of the inner edge, $v_1=1150\: \rm km\:s^{-1}$, and outer edge, $v_2=1700\: \rm km\:s^{-1}$, of the expanding shell \citep{2021ApJS..257...49C}, a constant electron temperature $T_e=10^4$\,K \citep{1974agn..book.....O}, and assumed pure hydrogen composition ($\rho$/$N_e$ = $1.67 \times 10^{-24}$\,g). The best fit ejecta mass was derived from this fit ($\chi^2=1234$, DOF=8) is $1.65\pm0.17\times 10^{-4}\: \rm M_\odot$. When compared to theoretical predictions by \citet{2005ApJ...623..398Y}, our derived ejecta mass places the mass of the WD that hosts V1369 Cen's explosion in the range 1--1.25\,$\rm M_\odot$. 

\subsubsection{YZ Ret}
YZ Ret is one of the brightest classical novae ($V_{peak}=3.7$\,mag), identified by R.~H.\ McNaught \citep{mcnaught} on 2020 July 15.590 UT at a distance of $2.7\substack{+0.4 \\ -0.3}$\,kpc \citep{2018AJ....156...58B}. It is associated with known cataclysmic variable MGAB-V207 and had not been reported as radio-loud until this search. 

Following previous multi-wavelength observations  \citep{2020ATel13867....1A,2020ATel13868....1L,2020ATel14214....1D,2020ATel14043....1S, 2020ATel14048....1I}, we measured a $\geq14\sigma$ detection for YZ Ret on 2022 Feb 6 at frequencies 2.9/5.5/9 GHz with the ATCA \citep{2022ATel15264....1G}. In this epoch, the peak radio flux density is somewhere between 2.9 and 9 GHz ($\alpha_{1.7-5.5\rm\,GHz} = 0.85\pm0.32$; $\alpha_{5.5-9.0\rm\,GHz} = -0.15$), which indicates that the emission is a mix of optically thick and optically thin thermal emission. Using relations derived by \citet{2021ApJS..257...49C}, we calculated the size of the radio photosphere to be $\ang{;;0}.65\pm\ang{;;0}.1$ and the corresponding surface brightness temperature to be $250\pm80$\,K. The surface brightness temperature was calculated at the peak of the 5\,GHz observations and an outer edge expanding ejecta velocity, $v_{2}$, of 2700$\,{\rm km\,s}^{-1}$ \citep{2020ATel13867....1A}. 
YZ Ret's surface brightness temperature is within the conservative temperature range for thermal radio emission ($\lesssim10^4$\,K), supporting the presence of optically thick thermal emission suggested in the spectral analysis of multi-frequency observations in this epoch. However, due to a lack of radio observations prior to our observations, this surface brightness temperature is not calculated at a known thermal maximum, and is insufficient to rule out the presence of non-thermal radio emission from YZ Ret. Thermal emission mechanisms are also suggested for YZ Ret in its study at X-ray wavelengths \citep[][]{2022MNRAS.514.2239S}.

For our 2022 Mar 10 follow-up observation, we measured a $\geq40\sigma$ detection at 5.5/9\,GHz and a $\geq7\sigma$ detection at 16.7/21.2\,GHz; for our 2022 Mar 18 observation, we measured a $\geq45\sigma$ detection at 5.5/9\,GHz, a $11\sigma$ detection at 16.7\,GHz, and no detection at 21.2\,GHz \citep{2022ATel15301....1G}. We measured steep spectral indices of $\alpha_{5.5-21.2\rm\,GHz}=-0.94\pm0.10$ and $\alpha_{5.5-16.7\rm\,GHz}= -1.00\pm0.15$ for the second and third epochs respectively. The fading of emission at higher frequencies, and the steep spectral index observed at lower frequencies indicates that free-free thermal emission component dominant at higher frequencies has decayed and only the synchrotron emission dominant at lower frequencies is being detected. This suggests that the ejecta are optically thin across the observed bands. Furthermore, the steepening of the spectral index from the second to third epoch suggests energy losses by the synchrotron \citep{1973ranp.book.....P}. These characteristics indicate that the classical nova YZ Ret was in its declining radio evolution phase at the time of our observations (see Fig. \ref{lightcurves} \textit{bottom panel left image}).

\subsubsection{RR Tel}
RR Tel is a symbiotic nova that evolves at an extremely slow rate \citep{1997A&A...323..387N} over a period of years to centuries. The analysis of X-ray observations for the nova by \citet{2013A&A...556A..85G} also suggest slow evolution, with the outflow from the WD being a direct result of the eruption even 65 years after the burst in 1944. The radio counterparts of RR Tel in VAST and RACS are offset by \ang{;;0}.6$-$\ang{;;1}.1 from its optical position. 

Over a year of ASKAP radio observations, RR Tel does not show any significant increase or decrease, as shown by the measured 0.888\,GHz flux densities ranging from $9$ to $10$\,mJy (see Fig. \ref{lightcurves} \textit{bottom panel right image}). Nonetheless, when compared to RR Tel detections with the ATCA in 2007 at a 8.6\,GHz flux density of $16.2\pm2.3 $\,mJy and a 4.8\,GHz flux density of $17.6\pm0.8$\,mJy \citep{2021ApJ...911...30D}, and previous radio observations at frequencies ranging from 5.0 to 8.9\,GHz \citep[]{1977ApJ...211..547S,1978MNRAS.184..893W,1982MNRAS.198..321P,1995MNRAS.273..517I, 2013ApJ...779...61M}, there is a visible trend of prolonged fading in the radio band.

\section{DETECTABILITY OF CLASSICAL NOVAE WITH ASKAP}
\label{s_discussion}

Our results from a targeted search demonstrated we can detect novae in a single epoch of the RACS survey, and in the VAST pilot surveys. In this section, we use previous novae radio studies in conjunction with ASKAP data to investigate the detectability of classical novae in ASKAP's full surveying mode (across a frequency range of 0.888--1.65\,GHz).

\begin{table}[t]
\begin{center}
\caption[Novae with simultaneous ASKAP and ATCA/VLA radio observations]{Summary of the properties of six slowly evolving radio classical novae from Sample 2 observed near-simultaneously with ASKAP and ATCA/VLA$^a$. Column 1 gives the GCVS ID of the nova that has simultaneous radio observations with ASKAP; Column 2 gives the number of days between discovery in optical and the ASKAP observation, $\Delta T$; Column 3 gives the number of days between ATCA or VLA observations at $t_{V}$ days post-discovery, and ASKAP observations at $t_{A}$ days post-discovery; Column 4 gives the observed ASKAP flux density, $S_{A}$; Column 5 gives the estimated flux density at 0.888\,GHz from the spectral fitting, $S_{est}$; Column 6 gives the spectral index, $\alpha$, for $\sim1-9$\,GHz; and Column 7 gives the nova evolution phase at the time of observation.}

\begin{adjustbox}{max width=\textwidth}
\begin{tabular}{rrrrrrrr}
\hline
\hline
\multicolumn{1}{c}{\text{GCVS ID}} & \multicolumn{1}{c}{\text{\begin{tabular}[c]{@{}c@{}}$\Delta$T \\ (days)\end{tabular}}} & \multicolumn{1}{c}{\text{\begin{tabular}[c]{@{}c@{}}$t_{V}-t_{A}$ \\ (days)\end{tabular}}}  & \multicolumn{1}{c}{\text{\begin{tabular}[c]{@{}c@{}}S$_{A}$ \\ (mJy)\end{tabular}}} & \multicolumn{1}{c}{\text{\begin{tabular}[c]{@{}c@{}}S$_{est}$ \\ (mJy)\end{tabular}}} & \multicolumn{1}{c}{\text{$\alpha$}} & \multicolumn{1}{c}{\text{\begin{tabular}[c]{@{}c@{}}Evolution \\ Phase\end{tabular}}} \\ \hline
V357 Mus  & 478 & 21.4  & $<0.53$ & $0.32\pm0.04$ & $1.27\pm0.10$ & Second Peak \\
V906 Car  & 413 & 25.4  & $<0.50$ & $0.29\pm0.07$ & $1.63\pm0.18$ & Rise \\
V5855 Sgr  & 917 & -26.5  & $<1.11$ & $0.19\pm0.01$ & $0.6\pm0.03$ & Decline \\
V5856 Sgr & 915 & -29.5  & $<0.78$ & $0.22\pm0.01$ & $1.26\pm0.03$ & Post Peak \\
V5667 Sgr & 1533 & -1.5 & $<0.94$ & $0.24\pm0.01$ & $1.08\pm0.03$ & Post Peak \\
V5668 Sgr & 1505 & -4.5 & $1.56\pm0.06$ & $1.47\pm0.36$ &   \multicolumn{1}{c}{\text{\begin{tabular}[c]{@{}c@{}}$1.65\pm0.55$; \\ $0.07\pm0.02^b$\end{tabular}}} & Decline \\ \hline
\vspace{-1cm}
\footnotetext{$^a$ATCA/VLA data available in \citet{2022yCat..22570049C}.\\
$^b$ The spectral indices for V5668 Sgr correspond to 1.26--1.74 and 1.74--7 GHz respectively.}
\end{tabular}
\end{adjustbox}
\label{table:5}
\end{center}
\end{table}

\begin{figure*}[t]
     \includegraphics[width=0.9\textwidth,trim={0cm 0cm 0cm 0cm},clip]{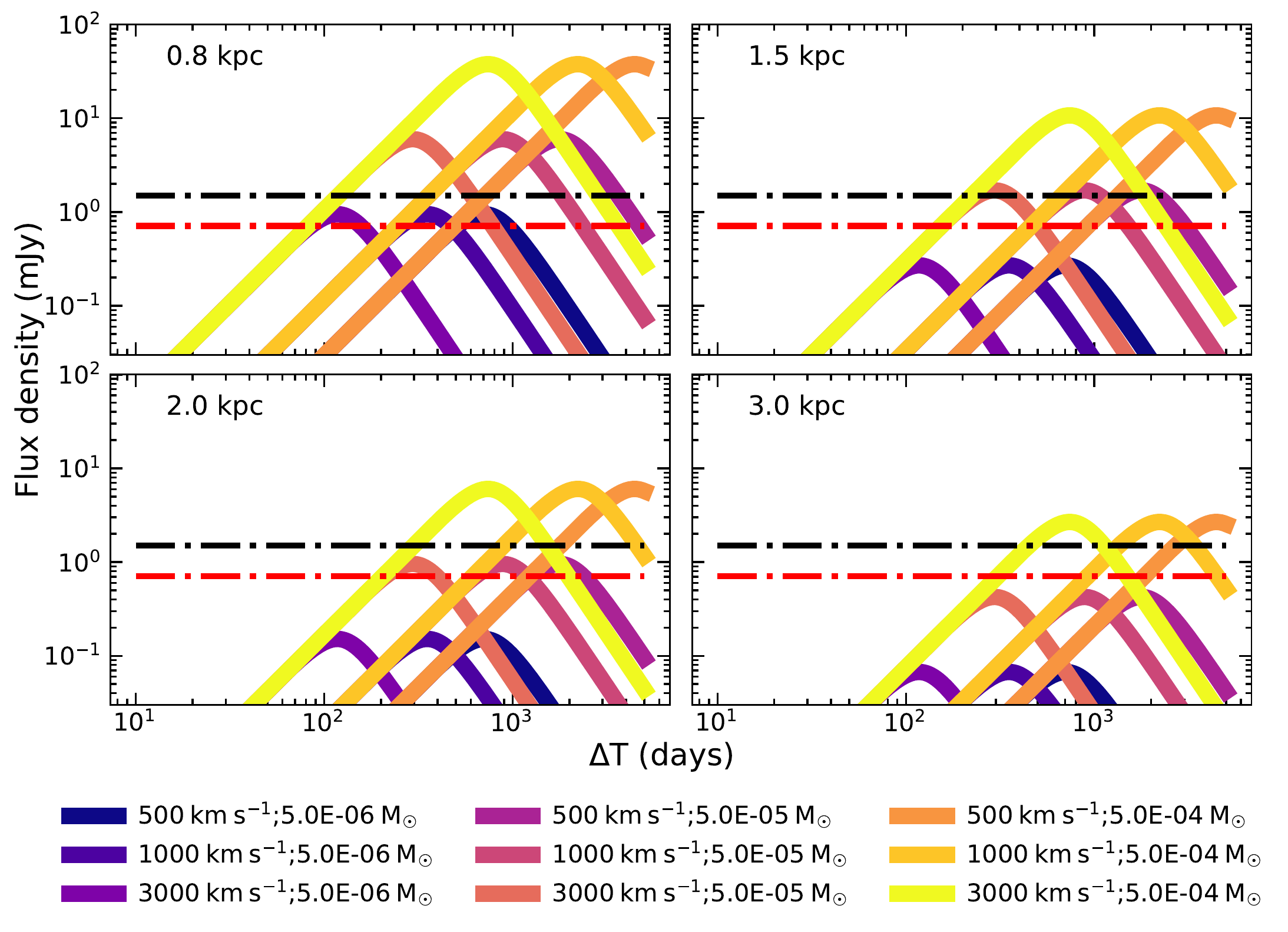}   
\caption{Simulated novae light curves at 1\,GHz using the Hubble Flow model with assumed pure hydrogen composition. The model used average parameters derived in literature: maximum ejecta velocity, $v_2$, from 500 to 3000 $\rm km\: s^{-1}$, ejected mass, $M_ej$, from $5\times10^{-6}$ to $5\times10^{-4}\: \rm M_\odot$, the ratio of minimum to maximum ejecta velocity, $v_1/v_2$, as 0.5, temperature of ejected material, $T_e$, as $10^4$\,K. The legend describes the combination of $v_2$ and $M_{ej}$ used for each curve. The nova distance is given in each plot. The VAST/RACS and the GASKAP 5$\sigma$ sensitivity limits are given by the black and red dotted-dashed lines respectively.}%
\label{simulatednovamodels}
\end{figure*}

We selected six novae from Sample 2 that had near-simultaneous ASKAP and ATCA/VLA multi-frequency observations. Five of the six novae detected in radio at higher frequencies -- V357 Mus, V906 Car, V5855 Sgr, V5856 Sgr, and V5667 Sgr -- were undetected with ASKAP at 0.888\,GHz (see Table \ref{table:5}). The spectral index scaling of near-simultaneous multi-frequency observations yields flux density estimates, $S_{est}$, in the range $\sim0.2-0.3$\,mJy at 0.888\,GHz, which are below ASKAP's VAST/RACS $5\sigma$ sensitivity threshold of 1.5 mJy (median RMS 0.326 mJy $\rm{beam^{-1}}$). Due to lack of flux-density sensitivity for the limited radio observations at ASKAP frequencies, we simulated radio lightcurves for novae at 1\,GHz using average parameters for ejected mass and velocity from \citet{2021ARA&A..59..391C} in the Hubble Flow model. These simulated curves are presented in Fig. \ref{simulatednovamodels} in relation with RACS/VAST and a deeper Galactic survey, GASKAP's 5$\sigma$ sensitivity limits.

As shown in the plots, the greater the mass ejected, the longer the nova eruption would be visible above the VAST or GASKAP 5$\sigma$ sensitivity limits. At 0.8 kpc, novae with $M_{ej}$ in the range $5\times 10^{-6}\rm \:M_\odot$ are not visible above the VAST detection limit. As the distance of the explosion increases to 3\,kpc, only explosions with $M_{ej}\gtrsim 5\times 10^{-4}\rm \:M_\odot$ are detectable above the VAST sensitivity limit.
With a 5$\sigma$ sensitivity limit of 1.5 mJy in 15 minute survey observations (RACS and VAST), we can detect radio emission up to a distance of 4 kpc if the ejecta mass is in the range $10^{-3}\rm \:M_\odot$, and upto 1 kpc if ejecta mass is in the range $10^{-5}-10^{-3}\rm \:M_\odot$. Since, the mass ejected is inversely proportional to the mass of the WD hosting the explosion, the detectability limit on the ejecta mass puts an upper limit of $1.25\:\rm M_\odot$ and $1\:\rm M_\odot$ on the masses of the white dwarfs that host ASKAP detectable novae explosions at 1 and 4 kpc respectively \citep{2005ApJ...623..398Y}. YZ Ret's inferred WD mass of $0.98\pm0.23\:\rm M_\odot$ in \citet{2022Natur.605..248K} at 2.7\,kpc is consistent with this ASKAP detectability limit. A higher 5$\sigma$ sensitivity limit of 0.7 mJy in the GASKAP survey, focused on the Galactic plane, will be useful for detecting radio emission from nova explosions hosted on higher mass host white dwarfs ($\gtrsim1\:\rm M_\odot$) that eject material $\lesssim5\times10^{-6}\:\rm M_\odot$. 

Fig. \ref{simulatednovamodels} also shows that as ejecta velocity, $v_2$, increases, the radio thermal maximum peaks earlier and spends less time above the 5$\sigma$ sensitivity limits (see Table \ref{simsimchar}). At 0.8 kpc, novae with $M_{ej}$ in the range $5\times 10^{-4}\rm \:M_\odot$ spend $\sim3000-4000$~days above the RACS/VAST 5$\sigma$ limit where the greater number of days correspond to lower velocities ($v_{2}\approx 500\: \rm km\:s^{-1}$). As the distance of explosion increases to 3\,kpc, for the same $M_{ej}$, the time spent above the RACS/VAST 5$\sigma$ limit is $\sim700-2000$ days. For faster novae ($v_{2}\approx 3000\: \rm km\:s^{-1}$), we would need one epoch of radio sky surveying per 2 months since the minimum time detectable above the higher sensitivity GASKAP 5$\sigma$ limit in the curves simulated for the lowest $M_{ej}\approx 5\times 10^{-6}\rm \:M_\odot$ is 72 days (see Table \ref{simsimchar}). VAST surveys, with their multi-epoch surveying strategy covering 265\,$\rm deg^2$ on the Galactic centre, would be the best ASKAP survey to capture fast novae. 
The Galactic centre has high dust concentration along its line of sight (see Fig. \ref{Fig:Distribution}), implying pronounced reddening and higher optical extinction \citep{2014A&A...571A..11P}. This region is, therefore, particularly suited for radio searches of novae with VAST.

Along with nova ejecta characteristics, and survey sensitivities, novae will of course only be detectable if they are within the sky coverage of a given survey. From the current pilot surveys listed in Table \ref{Surveys}, the three surveys that cover the highest percentage of all previous nova eruptions are RACS (93.6\%), VAST (31.9\%) and GASKAP (14.1\%). Single epoch ASKAP detections, from all-sky surveys such as RACS, will be useful for multi-frequency radio follow-up observations with other telescopes such as the ATCA, as demonstrated for the case of YZ Ret in Section \ref{s_results}. 

Because of the irregular cadence for pilot survey observations used in this work, multiwavelength data was required to determine radio counterparts. Future ASKAP surveys will increase the number of novae with radio lightcurves, allowing  spectral and temporal modeling. Radio detections could also overcome extinction effects and gaps in optical novae monitoring, which account for a large discrepancy of $\sim30-50$ novae between observed and theoretical discovery rates per year \citep{2017ApJ...834..196S,2021ApJ...912...19D}. For example, VVV-WIT-01 was classified as a classical nova using only radio and IR multiwavelength analysis, with no detection in optical \citep{2020MNRAS.492.4847L}. The availability of a statistically robust sample of novae observed in radio would also aid in more accurate ejecta mass predictions. While our targeted search did not uncover radio rebrightening from recurrent novae, data from routine ASKAP surveys will allow these experiments to enable insights about their ejecta masses and spectral signatures \citep[for e.g., RS Oph; ][]{2022ApJ...935...44C}.  

We have primarily considered the discovery rate of novae due to thermal emission, which can be faint at low radio frequencies ($\lesssim1$ GHz) even when not absorbed, as evidenced by the ASKAP non-detections for all novae in their optically thick thermal emission phase in Table \ref{table:5}. Furthermore, spectral index analysis for our detected novae in Section \ref{s_results} reveals that when the ejecta is optically thin, we identify non-thermal synchrotron emission for V1369 Cen and YZ Ret. This is consistent with previous observations of novae systems at low frequencies \citep{2012BASI...40..311K,2016MNRAS.457..887W,2023arXiv230110552D}. As long as the thermally-emitting material is not optically thick, synchrotron emission can be brighter at lower frequencies, and is present in 25\% of the radio novae examined in \citet{2021ApJS..257...49C}, with a hypothesised upper limit of 100\%. ASKAP observations of radio emission from novae at lower frequencies could help place strict limitations on the fraction of novae that generate synchrotron radiation. This will advance our understanding of the luminosity and timing of non-thermal components in novae explosions and aid in understanding how frequently these contribute to the observed diversity in radio lightcurves and, thereby, reflect the shock energetics of novae. Low-frequency VAST lightcurves, combined with high-frequency observations, could prove useful for studying different regions and physics of the nova system \citep[see review by][]{2012BASI...40..311K}.

\section{CONCLUSIONS}
\label{s_conc}
We have completed the first systematic search for radio counterparts of 440 classical novae previously identified in optical at centimetre wavelengths using RACS, VAST and other ASKAP pilot surveys. We identified three known radio classical novae, and one with no previous radio detection. The detection of V1369 Cen, V5668 Sgr, and YZ Ret, all in their declining optically thin phase, and RR Tel in its ongoing slow variability phase, demonstrates ASKAP's sensitivity to classical nova detection. We conducted ATCA observations of V1369 Cen and YZ Ret which indicate the presence of optically thin synchrotron component dominant at lower frequencies from the computed spectral indices. The multi-frequency observations also helped fit an ejecta mass of $1.65 \pm0.17\times 10^{-4}\: \rm M_\odot$ for V1369 Cen; and strengthen the hypothesis for optically thick thermal radio emission from YZ Ret in the first observed epoch.

The surveying nature of ASKAP observations eliminate the need for large-scale radio campaigns across decades as follow-up observations can be conducted for those detectable with ASKAP, for example, monitoring detections for novae with host WD masses in the range $0.4-1.25\: \rm M_\odot$, that erupt within a distance of 1\,kpc in 1.5\,mJy (5$\sigma$) flux density sensitivity surveys. Single-epoch detections of previously identified optical novae with observed false-positive rates of $4.8 \times 10^{-4}$, combined with multi-frequency follow-up strategy using other radio telescopes, could enable observations of novae near thermal radio maximum, as well as potential easier discovery of recurrent novae. Multi epoch VAST surveys with a region covering the optically extinct Galactic centre line of sight is well suited for radio searches of highly reddened novae with ASKAP. Detection and classification of the radio emission mechanism at ASKAP frequencies could aid the study of relative timing and luminosity of the components in the multi-component emission mechanism in novae and help understand underlying shock energetics.

The extension and application of this search method to surveys conducted at higher ASKAP frequencies of 1.36 and 1.65\,GHz with a more complete novae list that includes progenitors discovered in UV/IR searches \citep{2021ApJ...912...19D} and from space based optical telescopes \citep{2023MNRAS.518.5279M} will help better understand the scope of novae detectability with ASKAP. The position of the survey region on the sky, survey cadence, and survey flux density sensitivity are essential factors to consider when designing campaigns for targeting novae in the future. These observations will provide an opportunity to improve nova population statistics and study associated nova physics. To better understand the implications of these results, an unbiased search for a year's worth of ASKAP data can be conducted when the full surveys begin.

\begin{acknowledgement}
We thank the referee for their helpful comments. AG, JL and JP are supported by Australian Government Research Training Program Scholarships. DK is supported by NSF grant AST-1816492. RS acknowledges grant number 12073029
from the National Natural Science Foundation of China (NSFC). Parts of this research were conducted by the Australian Research Council Centre of Excellence for Gravitational Wave Discovery (OzGrav), project number CE170100004. 

This scientific work uses data obtained from Inyarrimanha Ilgari Bundara / the Murchison Radio-astronomy Observatory. We acknowledge the Wajarri Yamaji People as the Traditional Owners and native title holders of the Observatory site. CSIRO’s Australian Square Kilometre Array Pathfinder radio telescope is part of the Australia Telescope National Facility. Operation of ASKAP is funded by the Australian Government with support from the National Collaborative Research Infrastructure Strategy. ASKAP uses the resources of the Pawsey Supercomputing Research Centre. Establishment of ASKAP, Inyarrimanha Ilgari Bundara, the CSIRO Murchison Radio-astronomy Observatory and the Pawsey Supercomputing Research Centre are initiatives of the Australian Government, with support from the Government of Western Australia and the Science and Industry Endowment Fund. The Australia Telescope Compact Array is part of the Australia Telescope National Facility which is funded by the Australian Government for operation as a National Facility managed by CSIRO. We acknowledge the Gomeroi people as the Traditional Owners of the Observatory site.

This research has made use of the SIMBAD database \citep {2000A&AS..143....9W}, operated at CDS, Strasbourg, France, and \textit{Software:} \textsc{Astropy} \citep{2013A&A...558A..33A}, \textsc{MATPLOTLIB} \citep{2007CSE.....9...90H}, \textsc{NumPy} \citep{2020Natur.585..357H}, \textsc{VAST Tools} \citep{2021PASA...38...54M}, \textsc{MIRIAD}\citep{1995ASPC...77..433S}.

\end{acknowledgement}

\bibliography{vast1,vast2,rate_plot,bibliography}

\clearpage
\appendix
\section{Uncertain Nova Counterparts}
\label{Ruled-out Nova Counterparts}

In \S \ref{s_results}, we ruled two out of the six nova radio counterpart candidates as uncertain. For each candidate we provide an explanation; further observations would be required to confirm these as counterparts.

\textbf{V1280 Sco} is an extremely slow evolving classical nova that was discovered in an outburst on 2007 February 4.8 by \citet{2007IAUC.8803....1Y} at a distance of $1.1\pm0.5$\,kpc \citep{2009ATel.2063....1N}. Its matched radio source in RACS detected at a $6.5\sigma$ significance flux density of $1.62\pm0.04$\,mJy is \ang{;;3}.3 offset from its optical position at 4,469 days post-discovery ($t_\textrm{RACS}$). An ejecta expansion velocity of $\sim350\pm160$\,$\rm km\:s^{-1}$ \citep{2012A&A...545A..63C} gives an ejecta radius of $\ang{;;0}.8\pm\ang{;;0}.5$. The lack of previously confirmed radio observations, a single epoch detection in RACS, inconsistency of ejecta  radius with optical radio offset (\ang{;;2}.5 differential) make it difficult to assert whether V1280 Sco's radio counterpart is related to it, hence, we ruled its radio counterpart as uncertain.

\begin{figure}[]
    \centering
   \includegraphics[width=1\textwidth,trim={0.2cm 0.2cm 0.2cm 0.2cm},clip]{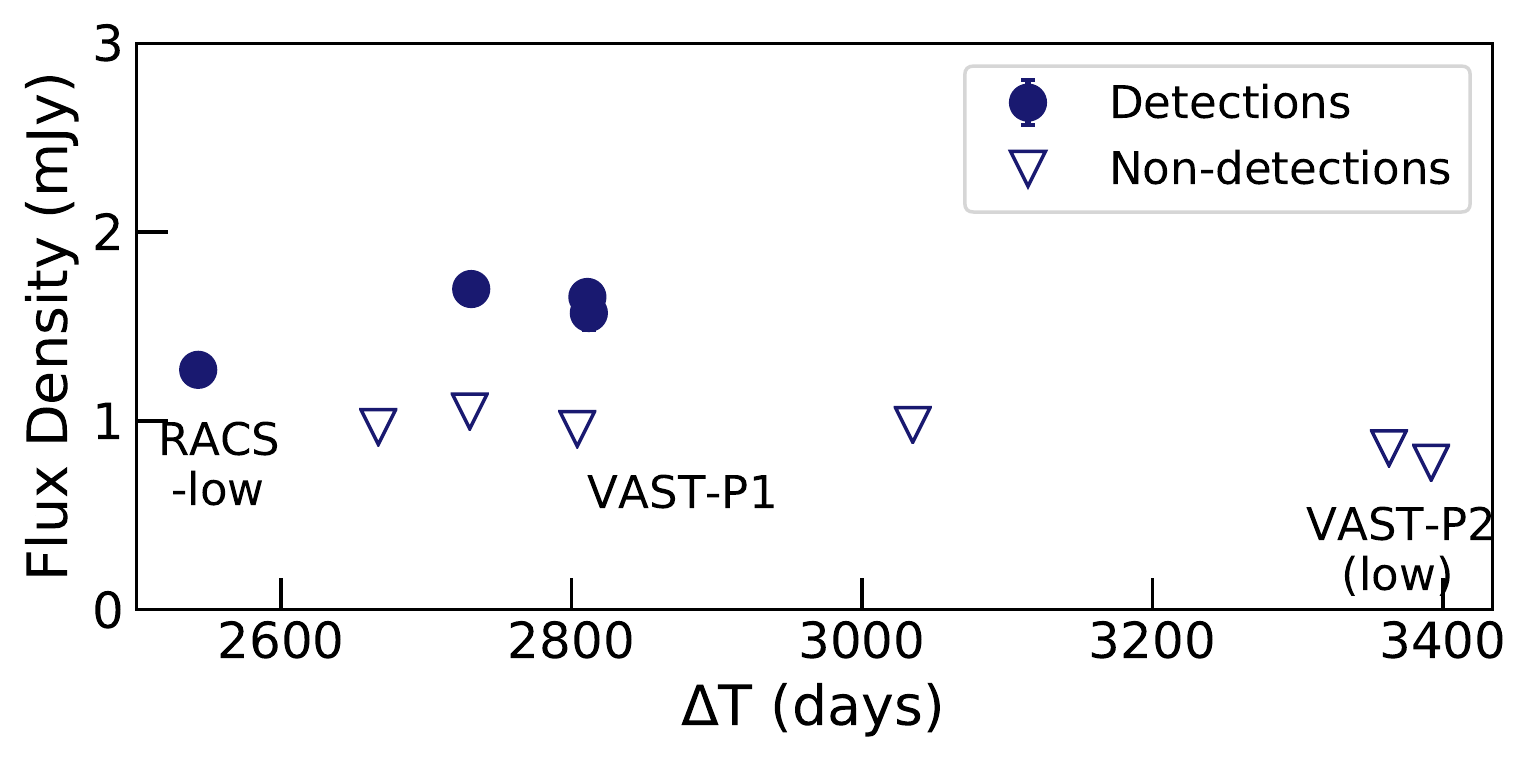}
\caption[]{The ASKAP radio lightcurve at V5590 Sgr's position. Filled-in circles indicate catalogued flux density within \ang{;;5} of the nova optical position, downward triangles are $3\sigma$ limits.}%
\label{fig:V5590 Sgr lc}%
\end{figure}

\textbf{V5590 Sgr} is a candidate Type D symbiotic nova candidate \citep{2014MNRAS.443..784M} announced in 2012 \citep{2012CBET.3140....2A,2012CBET.3140....1N}. Its matched radio source in VAST and RACS was not detected in all epochs of our search (see Fig \ref{fig:V5590 Sgr lc}). Despite its spatial consistency with the nova optical position (\ang{;;1}.5 offset), the near-detection limit 3$-$5$\sigma$ detections, lack of previous multiwavelength observations and confirmed radio detections make it difficult to assert
whether its radio counterpart is related to the nova or is some unrelated background source, hence, we ruled its radio counterpart as uncertain.

\newpage
\section{Simulated Radio Lightcurve Charcteristics}

In Table \ref{simsimchar}, we give the nova ejecta parameters from Fig \ref{simulatednovamodels} along with corresponding lightcurve characteristics.

\begin{table}[H]
\caption{Simulated nova radio lightcurve characteristics at 1\,GHz. Column 1 gives the ejected mass, $M_{ej}$; Column 2 gives the maximum ejecta velocity, $v_2$; Column 3 gives the maximum radio flux density, $S_{peak}$; Column 4 gives the time of $S_{peak}$, $t_{peak}$; Columns 5 and 6 give the time spent above VAST and GASKAP flux density sensitivity limits, $t_{VAST}$ and $t_{GASKAP}$.}
\begin{adjustbox}{width=1\textwidth}
\begin{tabular}{rrrrrrr}
\hline
\hline
\multicolumn{1}{c}{\begin{tabular}[c]{@{}c@{}}$ d$\\ (kpc)\end{tabular}} & \multicolumn{1}{c}{\begin{tabular}[c]{@{}c@{}}$ M_{ej}$\\ ($ \rm M_\odot$)\end{tabular}} & \multicolumn{1}{c}{\begin{tabular}[c]{@{}c@{}}$v_2$\\ ($\rm km\:s^{-1}$)\end{tabular}} & \multicolumn{1}{c}{\begin{tabular}[c]{@{}c@{}}$S_{peak}$\\ (mJy)\end{tabular}} & \multicolumn{1}{c}{\begin{tabular}[c]{@{}c@{}}$t_{peak}$\\ (days)\end{tabular}} & \multicolumn{1}{c}{\begin{tabular}[c]{@{}c@{}}$t_{VAST}$\\ (days)\end{tabular}} & \multicolumn{1}{c}{\begin{tabular}[c]{@{}c@{}}$t_{GASKAP}$\\ (days)\end{tabular}} \\
\hline

0.8 & 5e-6 & 500.0 & 0.95 & 704.0 & 0.0 & 438.0 \\
 & & 1000.0 & 0.95 & 352.0 & 0.0 & 219.0 \\
  & & 3000.0 & 0.95 & 117.0 & 0.0 & 72.0 \\
 & 5e-5 & 500.0 & 5.98 & 1768.0 & 2920.0 & 4230.0 \\
  & & 1000.0 & 5.98 & 884.0 & 1460.0 & 2115.0 \\
  & & 3000.0 & 5.98 & 294.0 & 486.0 & 705.0 \\
 & 5e-4 & 500.0 & 37.71 & 4440.0 & 4286.0 & 4512.0 \\
 &  & 1000.0 & 37.71 & 2220.0 & 4643.0 & 4756.0 \\
  & & 3000.0 & 37.71 & 740.0 & 2718.0 & 3578.0 \\
 \hline

1.5 & 5e-6 & 500.0 & 0.27 & 704.0 & 0.0 & 0.0 \\
& & 1000.0 & 0.27 & 352.0 & 0.0 & 0.0 \\
 & & 3000.0 & 0.27 & 117.0 & 0.0 & 0.0 \\
& 5e-5 & 500.0 & 1.7 & 1768.0 & 680.0 & 2122.0 \\
 & & 1000.0 & 1.7 & 884.0 & 340.0 & 1061.0 \\
 & & 3000.0 & 1.7 & 294.0 & 113.0 & 353.0 \\
& 5e-4 & 500.0 & 10.73 & 4440.0 & 3661.0 & 4085.0 \\
 & & 1000.0 & 10.73 & 2220.0 & 4330.0 & 4542.0 \\
 & & 3000.0 & 10.73 & 740.0 & 1636.0 & 2252.0 \\
\hline

2.0 & 5e-6 & 500.0 & 0.15 & 704.0 & 0.0 & 0.0 \\
&  & 1000.0 & 0.15 & 352.0 & 0.0 & 0.0 \\
 & & 3000.0 & 0.15 & 117.0 & 0.0 & 0.0 \\
& 5e-5 & 500.0 & 0.96 & 1768.0 & 0.0 & 1121.0 \\
 & & 1000.0 & 0.96 & 884.0 & 0.0 & 560.0 \\
 & & 3000.0 & 0.96 & 294.0 & 0.0 & 187.0 \\
& 5e-4 & 500.0 & 6.03 & 4440.0 & 3214.0 & 3780.0 \\
& & 1000.0 & 6.03 & 2220.0 & 3686.0 & 4390.0 \\
&  & 3000.0 & 6.03 & 740.0 & 1228.0 & 1778.0 \\
 \hline

3.0 & 5e-6 & 500.0 & 0.07 & 704.0 & 0.0 & 0.0 \\
 & & 1000.0 & 0.07 & 352.0 & 0.0 & 0.0 \\
 & & 3000.0 & 0.07 & 117.0 & 0.0 & 0.0 \\
& 5e-5 & 500.0 & 0.42 & 1768.0 & 0.0 & 0.0 \\
 & & 1000.0 & 0.42 & 884.0 & 0.0 & 0.0 \\
 & & 3000.0 & 0.42 & 294.0 & 0.0 & 0.0 \\
& 5e-4 & 500.0 & 2.68 & 4440.0 & 2299.0 & 3170.0 \\
 & & 1000.0 & 2.68 & 2220.0 & 2031.0 & 3587.0 \\
 & & 3000.0 & 2.68 & 740.0 & 677.0 & 1195.0\\
 \hline
\label{simsimchar}
\end{tabular}
\end{adjustbox}
\end{table}

\end{document}